\documentclass[aps,prb,showpacs,twocolumn,superscriptaddress]{revtex4}
\usepackage{amsmath,graphics}
\usepackage[next]{inputenc}
\usepackage[dvips]{epsfig}
\def\be{\begin{equation}}
\def\ee{\end{equation}}

\def\bi{\begin{itemize}}
\def\ei{\end{itemize}}
\def\bn{\begin{enumerate}}
\def\en{\end{enumerate}}
\def\bea{\begin{eqnarray}}
\def\eea{\end{eqnarray}}
\def\no{\nonumber}
\def\ba{\begin{array}}
\def\ea{\end{array}}
\def\bd{\begin{displaymath}}
\def\ed{\end{displaymath}}

\begin{document}
\title{The renormalization of entanglement in the anisotropic Heisenberg (XXZ) model}

\author{M. Kargarian}
\affiliation{Physics Department, Sharif
University of Technology, Tehran 11155-9161, Iran}
\author{R. Jafari}
\affiliation{Institute for Advanced Studies in Basic Sciences,
Zanjan 45195-1159, Iran}
\author{A. Langari}
\affiliation{Physics Department, Sharif University of Technology,
Tehran 11155-9161, Iran} \email[]{langari@sharif.edu}
\homepage[]{http://spin.cscm.ir}
\thanks{}
\altaffiliation{}

\begin{abstract}
We have applied our recent approach (Kargarian, \emph{et.al}
Phys. Rev. A {\bf 76}, 60304 (R) (2007))
to study the quantum information properties of
the anisotropic s=1/2 Heisenberg chain. We have investigated the
underlying quantum information properties like the evolution of
concurrence, entanglement entropy, nonanalytic behaviours and the
scaling close to the quantum critical point of the model. Both the
concurrence and the entanglement entropy develop two saturated
values after enough iterations of the renormalization of coupling constants.
This values are
associated with the two different phases, i.e N\'{e}el and spin
liquid phases. The nonanalytic behaviour comes from the divergence
of the first derivative of both measures of entanglement as the size
of system becomes large. The renormalization scheme demonstrates how
the minimum value of the first derivative and its position scales
with an exponent of the system size. It is shown that this exponent is
directly related to the critical properties of the model, i.e. the
exponent governing the divergence of the correlation length close to
the quantum critical point. We also use a renormalization method
based on the quantum group concept in order to get more insight
about the critical properties of the model and the renormalization of
entanglement.

\end{abstract}
\date{\today}

% insert suggested PACS numbers in braces on next line

\maketitle
%%%%%%%%%%%%%%%%%%%%%%%%%%%%%%%%%%%%%%%%%%%%%%%%%%%%%%%%%%%%%%%%%%%%%
\section{Introduction \label{introduction}}
The quantum information theory (QIT) and condensed matter physics
share in using the entanglement \cite{Bell}. The main motivation
behind such interest is two folds: On the one hand in  QIT the
entanglement is recognized as an essential resource for implementing
some quantum information tasks such as quantum computations, quantum
cryptography and densecoding \cite{Nielsen}. It is also believed
that the protocol based on the quantum entangled state has much
speed than the classical ones. On the other hand entanglement is a
unique measure of the quantum correlation of a pure state in
condensed matter physics. Thus, for condensed matter systems the
entanglement can de considered as some kind of correlation which
appears in the pure states.

The role of entanglement in quantum phase transition (QPT)
\cite{Sachdev} is of considerable interest\cite{Osterloh}. Quantum
phase transitions occur at absolute zero temperature. They are induced by the
change of an external parameter or coupling constant, and are driven
by quantum fluctuations. Quantum as well as classical phase
transitions are characterized by detecting nonanalytic behaviors in
some physical properties of the system. It is often accompanied by a
divergence in some correlation functions, but quantum systems
possess additional correlations which do not exist in a classical
counterparts, {\it the entanglement}. Entanglement is a direct
measure of quantum correlations and shows nonanalytic behavior such
as discontinuity in the vicinity of the quantum critical
point \cite{Wu,latorre}. In the past few years the subject of several
activities  were to investigate the behavior of entanglement in the
vicinity of quantum critical point for different spin models
\cite{Osterloh,Vidal1,Vidal2,Bose,Osborne,Verstraete} as well as
itinerant systems \cite{Zanardi,Gu,Anfossi}.

In our recent work \cite{kargarian} we have introduced the notion of
renormalization of entanglement and showed that this notion truly
describe the nonanalytic behavior of the derivative of entanglement
close to critical point of the Ising model in Transverse Field
(ITF). Moreover, we have investigated how the critical point is
reached by increasing the size of system via the renormalization
group (RG) approach. The finite size scaling demonstrates that the
RG of entanglement truly captures the critical behavior of the
model. The renormalization of quantum states has
also been introduced in terms of matrix product states \cite{fv}.

In this work we have applied the same approach to the
anisotropic spin 1/2 Heisenberg (XXZ) model in order to investigate
the critical behavior of the model using the evolution of
entanglement through the quantum renormalization group \cite{pfeuty,
wilson} close to quantum critical point.
%question
In this model the concurrence between the two nearest-neighbor sites
develops a maximum at the isotropic point ($\Delta=1$) without any
singularity in its first derivative \cite{Shi3}, i.e it vanishes at
the critical point $\Delta=1$. However, in our approach a
singularity, i.e the divergence of first derivative, is seen at the
critical point. It is expected that our approach is efficient and
powerful to study the quantum phase transition in various systems
since  the thermodynamic limit of the model is simply captured
through an analytic RG equations. Quantum renormalization group with
fixed boundary conditions give rise only to a nontrivial fixed point
$\Delta=1$ and trivial fixed points $\Delta=0,\infty$ for the XXZ
model while for every point in the $0\leq\Delta\leq1$ region the
model is critical, i.e gapless. A modification of the boundary conditions by
means of quantum group concept \cite{Mdelgado} restores
the gapless properties of $0\leq\Delta\leq1$, although the critical behavior of the model
and its scaling properties remain unchanged
from the gapped phas.

This article is organized as follows. In section-\ref{qrg} the main idea of
quantum renormalization group is briefly reviewed and in section-\ref{xxz-qrg}
this idea is applied to the XXZ model. Sections-\ref{ent-rg} discusses the
main idea of this paper about the renormalization of entanglement
entropy and concurrence is introduced and the section-\ref{qgqrg} is devoted
to the quantum group analysis in order to get more
insights about the critical features of the model. Finally in
section-\ref{conclusion} we summerize our results.

%%%%%%%%%%%%%%%%%%%%%%%%%%%%%%%%%%%%%%%%%%%%%%%%%%%%%%%%%%%%%%%%%%%%%%%%%%%%%

\section{Quantum renormalization group \label{qrg}}

The main idea of the RG method is the mode elimination or thinning
of the degrees of freedom followed by an iteration which reduces the
number of variables step by step until reaching a fixed point. In
Kadanoff's approach, the lattice is divided into blocks. Each block
is treated independently to build the projection operator onto the
lower energy subspace. The projection of the inter-block interaction
is mapped to an effective Hamiltonian ($H^{eff}$) which acts on the
renormalized subspace \cite{miguel1,langari}.
The procedure starts
by decomposing the Hamiltonian into two parts,
\bea \label{eq2}
H=H^{B}+ H^{BB}, \eea
where the block Hamiltonian $H^{B}$ is a sum
of commuting terms, each acting on different blocks. The
inter-block interaction is represented by  $H^{BB}$. The
perturbative implementation of this method has been discussed in
Refs.[\onlinecite{miguel1,jafari}]. We will shortly present this approach in
the 1st order correction. The zeroth order effective Hamiltonian is
given by
\bea \label{eq5} H^{eff}_{0}=P_{0}H^{B}P_{0}, \eea
 where $P_{0}$ is a projection operator. Since $H^{B}$ is a sum of
disconnected block Hamiltonians
\bea \no
H^{B}=\sum_{I=1}^{N'}h_{I}^{B}, \eea
one can search for a solution
of $P_{0}$ in a factorised form
\bea \no
P_{0}=\prod_{I=1}^{N'}P_{0}^{I}, \eea
where $N'$ is the number of
blocks. In the standard quantum renormalization group, $P_{0}^{I}$
is given by
\bea \no
P_{0}^{I}=\sum_{i=1}^{k}|\psi_{i}\rangle\langle\psi_{i}|, \eea
 where
$|\psi_{i}\rangle~(i=1,\cdots, k)$ are the $k$ lowest energy eigenstates
of $h_{I}^{B}$. The interaction between blocks define the 1st order
correction by the following equation \bea \label{eq6}
H_{1}^{eff}=P_{0}H^{BB}P_{0}. \eea The effective (renormalized)
Hamiltonian is then \be \label{eq7} H^{eff}=H^{eff}_{0}+H_{1}^{eff}.
\ee We will implement this approach in the next sections to obtain
the quantum properties of the  XXZ spin chain.

%%%%%%%%%%%%%%%%%%%%%%%%%%%%%%%%%%%%%%%%%%%%%%%%%%%%%%%%%%%%%%%%%
\section{Renormalization of the XXZ model \label{xxz-qrg}}
The Hamiltonian of XXZ model on a periodic chain of $N$ sites is
\bea
H(J,\Delta)=\frac{J}{4}(\sum_{i}^{N}\sigma_{i}^{x}\sigma_{i+1}^{x}+
\sigma_{i}^{y}\sigma_{i+1}^{y}+
\Delta\sigma_{i}^{z}\sigma_{i+1}^{z}) \label{eq23} \eea where
$J, \Delta>0$, $J$ is the exchange coupling, $\Delta$ is the
anisotropy parameter and $\sigma^{\alpha}$ ($\alpha=x, y z$) are
Pauli matrices. For $\Delta=1$, the Hamitonian is $SU(2)$ symmetry
invariant, but for $\Delta\neq1$ the $SU(2)$ symmetry breaks down
to the $U(1)$ rotational symmetry around the $z$-axis. The model is
exactly solvable by means of Bethe Ansats as far as the
rotational symmetry exists. It is known that for $0\leq\Delta\leq1$
the model is gapless with qasi-long range ordered where the
correlations decay algebraic with no magnetic long range order. For
$\Delta>1$, the symmetrty is reduced to $Z2$ and a gap opens which is
in the universality class of one dimensional antiferromagnetic Ising chain.
Indeed the third term in the Hamiltonian causes ordering in the
system and as $\Delta$ tends to infinity the N\'{e}el state is the
dominant phase of the system. The first two terms in the
Hamiltonian extend the quantum fluctuations in the system and
result in the corruption of the N\'{e}el ordering. It is shown that
the competition between the quantum fluctuations and ordering yield
a maximum value of concurrence between the two nearest neighbor sites
which show a scaling behavior at the critical point $\Delta=1$
\cite{Shi1},\cite{Shi2}.

To implement the  idea of QRG to calculate the entanglement
and concurrence we use a three site block procedure where the block
Hamiltonian is
$\\
\\h_{I}^{B}=\frac{J}{4}\Big[(\sigma_{1,I}^{x}\sigma_{2,I}^{x}+\sigma_{2,I}^{x}\sigma_{3,I}^{x}+
\sigma_{1,I}^{y}\sigma_{2,I}^{y}+\sigma_{2,I}^{y}\sigma_{3,I}^{y})+$
\bea\label{eq24}
\Delta(\sigma_{1,I}^{z}\sigma_{2,I}^{z}+\sigma_{2,I}^{z}\sigma_{3,I}^{z})\Big].
\eea In this case the inter-block ($H^{BB})$ and intra-block
($H^{B}$) Hamiltonians are \bea \label{eq25}
H^{BB}=\frac{J}{4}\left[\sum_{I}^{N/3}(\sigma_{3,I}^{x}\sigma_{1,I+1}^{x}+
\sigma_{3,I}^{y}\sigma_{1,I+1}^{y}+\Delta\sigma_{3,I}^{z}\sigma_{1,I+1}^{z})\right],
\eea \bea \label{eq26} H^{B}=\sum_{I}^{N/3} h_{I}^{B}.
 \eea
A remark is in order here, choosing the three site block is
essential here to get a self similar Hamiltonian after each RG
step. An odd site XXZ Hamiltonian has two degenerate ground states
which are used to construct the projection operator of quantum RG (QRG).
These
degenerate ground states are \bea \label{eq27-1}
|\phi_{0}\rangle=\frac{1}{\sqrt{2+q^{2}}}(|\uparrow\uparrow\downarrow\rangle+
q|\uparrow\downarrow\uparrow\rangle+|\downarrow\uparrow\uparrow\rangle),\eea
\bea \label{eq27-2}
|\phi_{0}'\rangle=\frac{1}{\sqrt{2+q^{2}}}(|\uparrow\downarrow\downarrow\rangle+
q|\downarrow\uparrow\downarrow\rangle+|\downarrow\downarrow\uparrow\rangle),
\eea
where $q$ is \be \label{eq28}
q=\frac{-1}{2}[\Delta+\sqrt{\Delta^{2}+8}]. \ee
The corresponding
energy is \bea \label{eq29}
e_{0}=-\frac{J}{4}[\Delta+\sqrt{\Delta^{2}+8}], \eea and $|\uparrow
\rangle$, $|\downarrow \rangle$ are the eigenstates of $\sigma^{z}$.

\begin{figure}
\begin{center}
\includegraphics[width=8cm]{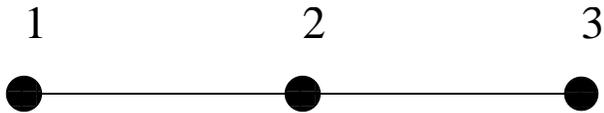}
\caption{Three sites in a block labeled sequentially.}
\label{3siteblock}
\end{center}
\end{figure}

The  projection operator ($P_{0}$) for the $I$-th block is defined
\bea \label{eq30}
P_{0}^{I}={|\Uparrow\rangle}_{II}\langle\phi_{0}|+{|\Downarrow\rangle}_{II}\langle\phi_{0}'|,
\eea where $|\Uparrow\rangle_{I}$ and $|\Downarrow \rangle_{I}$ are
renamed states of each block to represent the effective site
degrees of freedom. The renormalization of Pauli matrices are given
by \bea \label{eq31}
P_{0}^{I}\sigma_{i,I}^{\alpha}P_{0}^{I}=\xi_{i}^{\alpha}{\sigma'}_{I}^{\alpha}~~~~~~(i=1,2,3~~;~~\alpha=x,y,z),
\eea where
\bea
\label{eq32}
\xi_{1}^{x,y}=\xi_{3}^{x,y}=\frac{2q}{2+q^{2}}~~~&,&~~~
\xi_{1}^{z}=\xi_{3}^{z}=\frac{q^{2}}{2+q^{2}} \nonumber \\
%\eea \bea \label{eq32}
\xi_{2}^{x,y}=\frac{2}{2+q^{2}}~~~&,&~~~
\xi_{2}^{z}=\frac{2-q^{2}}{2+q^{2}}, \eea where the indices 1, 2 and
3 refer to sites labeled in a single block based on
Fig.(\ref{3siteblock}).

%%%%%%%%%%%%%%%%%%%%%  Fig.20   %%%%%%%%%%%%%%%%%%%%%%%

%%%%%%%%%%%%%%%%%%%%%%%%%%%%%%%%%%%%%%%%%%%%%%%%%%%%%%%
 The effective Hamiltonian (using
Eq.(\ref{eq7})) of the renormalized chain is again an XXZ chain with
the scaled couplings \bea \label{eq33}
H^{eff}=\frac{J'}{4}\left[\sum_{i}^{N/3}({\sigma}_{i}^{x}{\sigma}_{i+1}^{x}+{\sigma}_{i}^{y}{\sigma}_{i+1}^{y})+
\Delta'({\sigma}_{i}^{z}{\sigma}_{i+1}^{z})\right],\eea where \bea
\label{eq34}
J'=J(\frac{2q}{2+q^{2}})^{2}~~~,~~~\Delta'=\Delta\frac{q^{2}}{4}.
\eea

%{\bf A short discussion on the QRG flow should appear here.}

The stable and unstable fixed points of the QRG equations is
obtained by solving $\Delta'=\Delta\equiv\Delta^*$. The stable fixed points locate
at zero and infinity while $\Delta^*=1$ stands for the unstable fixed
point which specify the critical point of the model. Starting with
any initial value for $\Delta>1$, the coupling constant flows toward
infinity showing that the model falls into the universality class of
Ising model, while for $\Delta<1$ the stable $\Delta=0$ fixed point
is touched. As  we have mentioned previously, for $0\leq\Delta\leq1$
the model represents a spin fluid phase. The transition between the
two phases are truly captured by real space QRG. The main
discrepancy of our results on the Ising model in transverse field \cite{kargarian} (ITF)
and the present one on XXZ
 comes from the fact that  XXZ
model  is critical for all values of $0\leq\Delta\leq1$ but  QRG
equations do not to show the whole critical region except $\Delta=0$. In
fact the QRG prescription only represents the masslessness property
of the XY fixed point ($\Delta=0$). However, when the coarse graining
procedure implemented by some appropriate boundary conditions in
order to get more correlation between blocks as is done in the
quantum group method \cite{Mdelgado}, the critical line of the model
is truly predicted (see section \ref{qgqrg}).

\section{Renormalized entanglement and its scaling property \label{ent-rg}}
%\section{mehdi-scaling in the ITF model}
\subsection{\emph{Entanglement and concurrence}}
In this subsection we calculate the concurrence and entanglement of
the XXZ chain using our proposal which implements the idea of
renormalization group. As we have mentioned previously, a finite
size block is treated exactly to calculate the physical quantities.
The coupling constants of a finite block are renormalized via the
QRG prescription to give the large size behaviour. The XXZ spin 1/2
Hamiltonian has two degenerate ground state in the three site block,
however to define the density matrix we have to consider one of
them.
%Otherwise, if we use a density matrix of mixed states the resulting
%entanglement does not specify the property of a single state.
Thus, the density matrix is defined by
\be \label{eq35}
\varrho=|\phi_{0}\rangle\langle\phi_{0}|, \ee
where
$|\phi_{0}\rangle$ has been introduced in Eq.(\ref{eq27-1}). The
results will be the same if we consider $|\phi'_{0}\rangle$ to
construct the density matrix.

\begin{figure}
\begin{center}
\includegraphics[width=8cm]{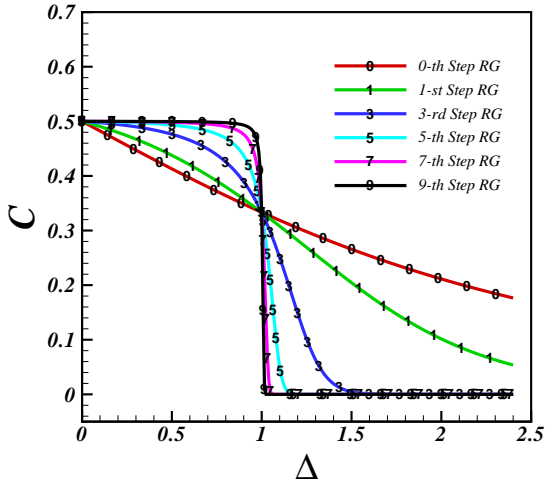}
\caption{(color online) Representation of the evolution of
concurrence in terms of RG iterations (steps).} \label{fig22}
\end{center}
\end{figure}

There are basically two choices to define the concurrence and
entanglement for a three site block. ({\it 1}) The symmetric case,
in which the concurrence between site 1 and 3 is obtained by summing
over the degrees of freedom of site 2. And the entanglement
between site 2 and remaining sites of the block as measured by
von-neuman entropy. ({\it 2}) We sum over site 1 or 3 and get the
concurrence between the middle site (2) and one site at the corner
side of the block (see Fig.(\ref{3siteblock})). Without loss of
generality we only concentrate on case ({\it 1}). The density matrix
defined in Eq.(\ref{eq35}) is traced over site 2 degrees of freedom
to get the reduced density matrix for site 1 and 3 ($\varrho_{1,3}$)
which gives \bea \label{eq36} \varrho_{13}=\frac{1}{2+q^{2}} \left(
  \begin{array}{cccc}
    q^{2} & 0 & 0 & 0 \\
    0 & 1 & 1 & 0 \\
    0 & 1 & 1 & 0 \\
    0 & 0 & 0 & 0 \\
  \end{array}
\right). \eea
The corresponding eigenvalues of
$\hat{R}=\varrho_{13}\tilde{\varrho_{13}}$ (where
$\tilde{\varrho}_{13}=
(\sigma^{y}_{1}\otimes\sigma^{y}_{3})\varrho_{13}^{\ast}(\sigma^{y}_{1}\otimes\sigma^{y}_{3})$)
are in ascending order
\bea \label{eq37} \lambda_{1}= \lambda_{2}=
\lambda_{3}=0\;\;,\;\; \lambda_{4}=\frac{4}{(2+q^{2})^{2}}. \eea
Thus, the
concurrence is obtained \bea \label{eq38}
C_{13}=\max\lbrace\lambda_{4}^{1/2}, 0\rbrace=\frac{2}{2+q^{2}}.
\eea
According to Eq.(\ref{eq28}), $C_{13}$ is a function of
$\Delta$. The renormalization of $\Delta$ defines the evolution of
concurrence as the size of system becomes large. We have plotted in
FIG.\ref{fig22} the value of $C_{13}$ versus $\Delta$ for different
QRG iterations.
%%%%%%%%%%%%%%%%%%%%%  Fig.22   %%%%%%%%%%%%%%%%%%%%%%%

\begin{figure}
\begin{center}
\includegraphics[width=8cm]{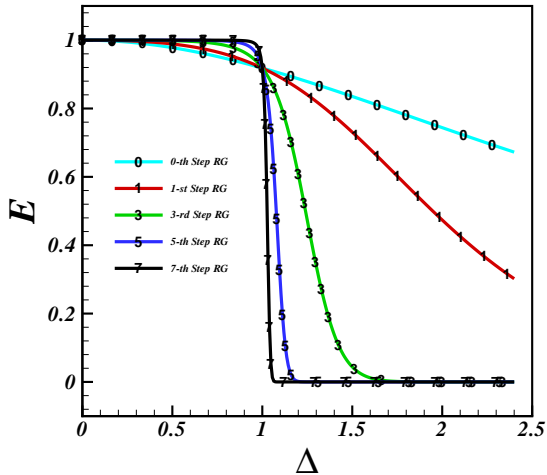}
\caption{(color online) The evolution of Entanglement entropy as the
size of system grows (i.e. different RG iterations).} \label{fig21}
\end{center}
\end{figure}

%%%%%%%%%%%%%%%%%%%%%%%%%%%%%%%%%%%%%%%%%%%%%%%%%%%%%%%
The plots of $C_{13}$ versus $\Delta$ for different QRG steps cross each other at the scale
invariant critical point, $\Delta_c\equiv\Delta^*=1$. By increasing the size of
system (higher QRG steps) the concurrence develops two
different behaviours which are separated at $\Delta_c$. After enough
QRG steps the value of $C_{13}$ is saturated for $0\leq\Delta<1$
which shows the existence of quantum correlations between two blocks
at large distances in an infinite chain that is effectively
described by a three site model with the renormalized coupling
constants. In this region ($0\leq\Delta\leq1$) the quantum fluctuations
arising from the transverse interactions have dominant effect and
destroy any long range order. In spite of lacking a
long range order, the evolution of concurrence via RG indicates that
the  spin fluid phase contains the quantum correlations, i.e the
qubits in the presence of quantum fluctuations are quantum
correlated. For $\Delta>1$, $C_{13}$ saturates the \emph{zero} value
asymptotically to represent the lack of quantum correlation in the
Ising limit. We have also examined the entanglement of formation
\cite{entanglement of formation} between sites 1 and 3 which can be
obtained by the following relation: \bea \label{eq39}
E_1&=&-y\log_{2}(y)-(1-y)\log_{2}(1-y), \no \\
y&=&\frac{1}{2}+\frac{1}{2}\sqrt{1-C_{13}^{2}}. \eea
The
entanglement of formation shows similar behavior to
FIG.\ref{fig22} because of monotonous relation between concurrence
and entanglement of formation. In the symmetric case we can also study
the entanglement between the middle site (2) with the remaining
sites of the block. The amount of entanglement is given by von-Neuman
entropy of reduced density matrix obtained after tracing out the
remaining sites of block
\bea \label{eq43}
 \varrho_{2}=\frac{1}{2+q^{2}}
\left(
\begin{array}{cc}
2 & 0 \\
0 & q^{2} \\
\end{array}
\right).\eea The von-Neuman entropy which is the entanglement of
site 2 is \bea \label{eq44}
E=-\frac{2}{2+q^{2}}\log_{2}\frac{2}{2+q^{2}}-
\frac{q^{2}}{2+q^{2}}\log_{2}\frac{q^{2}}{2+q^{2}}. \eea The
variation of entanglement ($E$)  versus $\Delta$ has been plotted in
Fig.\ref{fig21}. Different plots show the evolution of $E$ under QRG
iterations. In other words, the different step of QRG show how the
entanglement evolves when the size of chain is increased. Similar to
$C_{13}$, $E$ behaves as an order parameter which gets a nonzero
value for $0\leq\Delta<1$ and zero for $\Delta>1$ in the infinite
size limit. \emph{Nonzero} $E$ for $0\leq\Delta<1$ verfies again
that the state of model is entangled for $0\leq\Delta<1$ where the
ground state is characterized by a gapless excitation and algebraic
decay of spin correlations (the spin fluid phase). While, $E$ is
zero for the N\'{e}el state ($\Delta>1$) which is not an entangled
state.
%%%%%%%%%%%%%%%%%%%%%  Fig.21   %%%%%%%%%%%%%%%%%%%%%%%

%%%%%%%%%%%%%%%%%%%%%%%%%%%%%%%%%%%%%%%%%%%%%%%%%%%%%%%

As previously discussed the entanglement between the middle site (2)
and the remaining ones tends to maximal entanglement for the region
of coupling $0\leq \Delta<1$ after few RG steps Fig.\ref{fig21}.
While the entanglement of formation between sites 1 and 3 tends to a
nonmaximal value, in fact this is a monogamy property of
entanglement\cite{Coffman}. On the basis of monogamy property , the
maximal entanglement between two parties restricts their
entanglement with third party and vis versa.

\subsection{\emph{nonanalytic and scaling behavior}}
The non-analytic behavior in some physical quantity is a feature of
second-order quantum phase transition. It is also accompanied by a
scaling behavior since the correlation length diverges and there is
no characteristic length scale in the system at the critical point.
There have been a number of theoretical studies on the entanglement
and quantum phase transitions. These studies show that the
entanglement as a direct measure of quantum correlations indicate
critical behavior such as diverging of its derivative  as the phase
transition is crossed. Osterloh,\emph{et. al}\cite{Osterloh} have
verified that the entanglement in the vicinity of critical point of
ITF and XX model in transverse field show a scaling behavior. It has
also been shown that the entanglement between a block of spins and
the rest of system scales near the quantum critical
point \cite{Vidal1}. The scaling properties of entanglement with the
size of block differs at and away from the critical point. At the
critical point where the correlations decay algebraically entanglement scales
logarithmically while it saturates away from the critical point
\cite{Latorre}. It can be interpreted in the framework of conformal
field theory \cite{korepin} associated with the quantum phase
transition and the central charge of the theory. Indeed for the XXZ
model, the entanglement entropy, i.e. the entanglement of a block
(sublattice bipartition) with the rest of the system, demonstrates
extermum behavior at the critical point\cite{Chen}.

In this work, we
adopt a preferable and distinct way to study the block entanglement
via the renormalization group approach. As we have stated in the RG
approach for XXZ model, a large system, i.e. $N=3^{n+1}$, can be
effectively described by three sites with the renormalized couplings
of the n-th RG iteration. Thus, the entanglement between the two
renormalized sites represents the entanglement between two parts of
the system each containing $N/3$ sites effectively. In this respect
we can speak of {\it block entanglement} - block-block entanglement
or the entanglement between a block and the rest of system. It is
shown that the first derivative of the entanglement shows a
diverging behavior as the critical point is reached.

\begin{figure}
\begin{center}
\includegraphics[width=8cm]{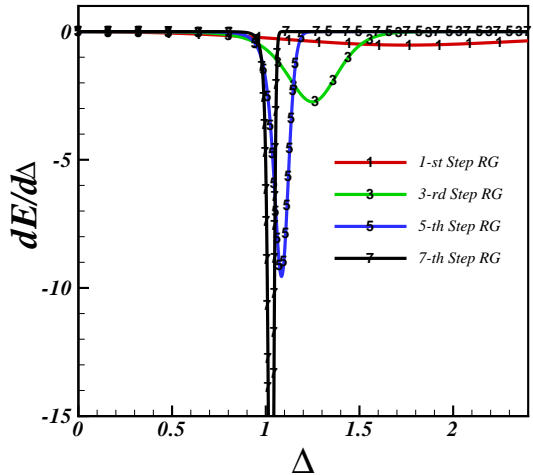}
\caption{(color online) First derivative of entanglement entropy and
its manifestation towards diverging as the number of RG iterations
(steps) increases (Fig.\ref{fig21}).} \label{fig34}
\end{center}
\end{figure}

\begin{figure}
\begin{center}
\includegraphics[width=8cm]{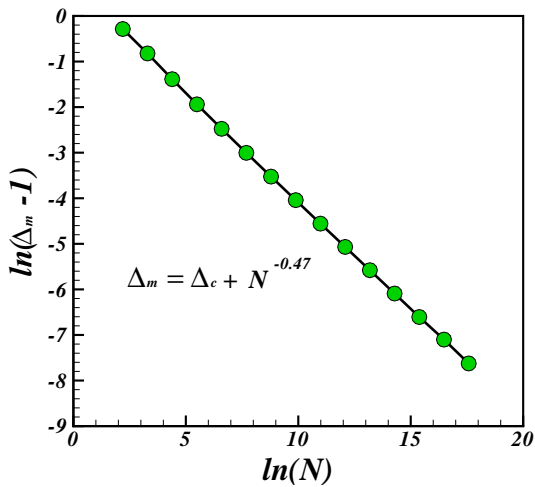}
\caption{(color online) The scaling behavior of $\Delta_{m}$ in
terms of system size ($N$) where $\Delta_{m}$ is the position of
minimum in Fig.\ref{fig34}.} \label{fig35}
\end{center}
\end{figure}

\begin{figure}
\begin{center}
\includegraphics[width=8cm]{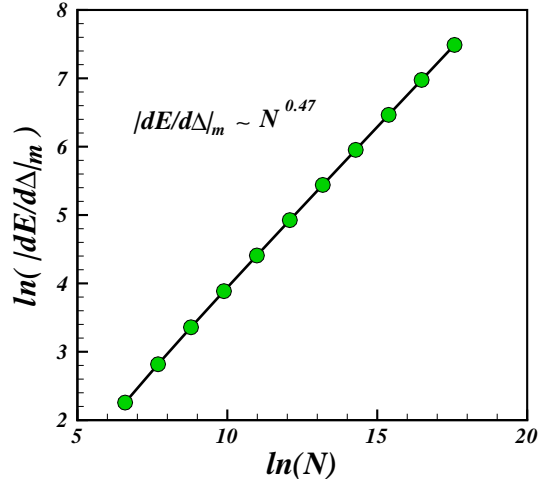}
\caption{(color online) The logarithm of the absolute value of
minimum, $\ln(\mid dE/d\Delta\mid_{\Delta_{m}})$, versus the
logarithm of chain size, $\ln(N)$, which is linear and shows a
scaling behavior. Each point corresponds to the minimum value of a
single plot of Fig.\ref{fig34}. } \label{fig36}
\end{center}
\end{figure}

Having this in mind, the first derivative of concurrence is analyzed
as a function of coupling $\Delta$ at different RG steps which
manifest the size of system. The derivative of concurrence with
respect to the coupling constant ($\frac{dF}{dg}$) shows a singular
behavior at the critical point. It is given by \be \label{eq20}
\frac{dF^{(n)}}{d\Delta}=\frac{dF^{(n)}}{d\Delta_{n}}\frac{d\Delta_{n}}{d\Delta_{n-1}}{...}\frac{d\Delta_{1}}{d\Delta}
%\frac{dC}{dg}=\Big( -\frac{1}{\ln2}\big(\frac{2\theta}{(1+\theta^{2})^{2}}\ln
%\theta^{2}\big) \Big)\times \Big(2-\frac{4g}{\sqrt{4g^{2}+1}}\Big).
\ee
 where $F^{(n)}$   stand for the concurrence or von-Neuman entropy and $\Delta_{n}$
the renormalized anisotropy coupling ($\Delta$) at the n-th RG
iteration. The singular behavior is the result of discontinuous
change of $C$ at $\Delta=\Delta_c$. We only concentrate on the
nonanalytic behavior of entanglement between the middle site and
remaining sites of each block. We have plotted $\frac{dE}{d\Delta}$
versus $\Delta$ in Fig.\ref{fig34} for different RG iterations which
shows the singular behavior as the size of system becomes large
(higher RG steps). A more detailed analysis shows that the position
of the minimum ($\Delta_m$) of $\frac{dE}{d\Delta}$ tends towards
the critical point like $\Delta_{m}=\Delta_{c}+N^{-0.47}$ which has
been plotted in Fig.\ref{fig35}. Moreover, we have derived the
scaling behavior of $y\equiv|\frac{dE}{d\Delta}|_{\Delta_m}$ versus
$N$. This has been plotted in Fig.\ref{fig36} which shows a linear
behavior of $ln(y)$ versus $ln(N)$. The exponent for this behavior
is $\mid\frac{dE}{d\Delta}|_{\Delta_m} \sim N^{0.47}$. This results
justify that the RG implementation of entanglement  truly capture
the critical behavior of the XXZ model at $\Delta=1$.
%%%%%%%%%%%%%%%%%%%%%  Fig. first derivative entanglement(2-13)  %%%%%%%%%%%%%%%%%%%%%%%
%%%%%%%%%%%%%%%%%%%%%%%%%%%%%%%%%%%%%%%%%%%%%%%%%%%%%%%%%%%%

%%%%%%%%%%%%%%%%%%%%%  Fig. scaling of position $\Delta_{min}$   %%%%%%%%%%%%%%%%%%%%%%%
%%%%%%%%%%%%%%%%%%%%%%%%%%%%%%%%%%%%%%%%%%%%%%%%%%%%%%%%%%%%

%%%%%%%%%%%%%%%%  Fig.power scaling of maximum of derivative of entanglement(2-13)  %%%%%%%%%%%%%%%%%%%%%%%

%%%%%%%%%%%%%%%%%%%%%%%%%%%%%%%%%%%%%%%%%%%%%%%%%%%%%%%
Besides the entanglement entropy, two point quantum correlation as
measured by concurrence develops a maximum at the critical point.
The concurrence of the two nearest neighbor sites of a large chain
has numerically been calculated and the maximum at the critical
point is due to the counterbalance of ordering and quantum
fluctuations \cite{Shi3}. In the Ising limit, i.e $\Delta
\longrightarrow \infty$, the N\'{e}el order is developed which makes
the concurrence becomes zero. The same treatment also is seen in the
renormalization of entanglement entropy or concurrence. As stated in
section (2) the renormalisation equations develops two different phases
for the XXZ model which are separated by the unstable fixed point $\Delta_c
= 1$. For $\Delta > 1$ region which falls into the Ising
universality class, both the entanglement entropy and concurrence
(Fig.\ref{fig22}) tends to zero as the iteration of RG equations
increases, i.e the stable fixed point $\Delta \longrightarrow
\infty$ is reached.
%given more insight near the phase transition, in
%addition the  entanglement entropy, the first derivative of
%concurrence and scaling are examined.

A novel issue arises when the exponents of the scaling of
$\Delta_{m}$ and the minimum value of von-Numan entropy, as
discussed in the previous paragraphs, are compared with the
corresponding ones of the concurrence. The first derivative of
concurrence tends to diverge at the quantum critical point. The
position of minimum and the minimum value itself scale as
$\Delta_{m}=\Delta_{c}+N^{-0.46}$ and
$\mid\frac{dC}{d\Delta}|_{\Delta_m} \sim N^{0.48}$, respectively, as
the size of the system becomes large. It is clear that the exponents
are almost the same for the two measures of entanglement. It
indicates that the entanglement entropy and concurrence scale in the
same manner in the limit of large scale behavior and are associated
to the long range behavior of the model close to the critical point.
According to this intuitive picture of entanglement scaling and
critical properties of the model we would like to emphasize that the
exponent is directly related to the correlation length exponent
close to the critical point. The correlation length exponent, $\nu$,
gives the behavior of correlation length in the vicinity of
$\Delta_{c}$ \be \label{eq50} \xi \sim (\Delta-\Delta_{c})^{-\nu}
\ee
Under the RG transformations (Eq.(\ref{eq34})) the correlation
length scales as $\xi \longrightarrow \xi^{(1)}=\xi/n_{B}$ where
$n_{B}$ stands for the number of sites in each block which is
$n_{B}=3$ in our procedure. So, it is easy to look for the
$n^{th}$ RG iteration
\be \label{eq51}\xi^{(n)} \sim
(\Delta_{n}-\Delta_{c})^{-\nu}=\xi/n_{B}^{n} \ee which immediately
leads to an expression for
$\mid\frac{d\Delta_{n}}{d\Delta}\mid_{\Delta_{c}}$ in terms of $\nu$
and $n_{B}$ \be
\label{eq52}\mid\frac{d\Delta_{n}}{d\Delta}\mid_{\Delta_{c}}\sim
N^{1/\nu}. \ee The comparison with Eq.(\ref{eq20}) demonstrates that
the exponent which governs the nonanalytic behavior of the
entanglement entropy and concurrence in the vicinity of the critical
point is nothing than the inverse of correlation length exponent. It
should be noted that the scaling of the position of minimum,
$\Delta_{m}$ (Fig.\ref{fig35}), also comes from the behavior of
the correlation length near the critical point. As the critical
point is approached and in the limit of large system sizes, not the
thermodynamic limit, the correlation length almost covers the whole
of  system, i.e. $\xi \sim N $, and a simple comparison with
Eq.(\ref{eq50}) results in the scaling of position as
$\Delta_{m}=\Delta_{c}+N^{-1/\nu}$.

\section{further insight from quantum group \label{qgqrg}}
As it has been carried out in Sec.(\ref{qrg}, \ref{xxz-qrg}) the (block) quantum
renormalization group is conceptually and technically
simple, but it may yield poor quantitative results. This was the main
reason on the slow development of QRG through the 1980s in favor of
powerfull numerical Quantum Mont Carlo method. However there was a
comback to the RG in the 1990s as one of the most powerful method
when dealing with the zero temperature properties of the quantum systems.
It was Wilson the first to associate the failure of the QRG
to the role of the boundary conditions when
applying to the tight bonding model.
In fact the success of the density matrix renormalization group (DMRG)
as developed by White\cite{white}
refers to the way in which it takes into account
the effect of boundary conditions in terms of
the quantum correlation in the ground sate of the system.
Later, it was shown that the
breakdown of the QRG is rooted on taking into account the
entanglement in the ground state of the system \cite{JOsborne}.

However, Martin-Delgado, \emph{et
al}\cite{Mdelgado} proposed a QRG prescription which implements the
concept of the quantum group to the renormalization group approach.
In this approach the effect of boundary conditions have been imposed in
terms of boundary magnetic fields on each block. The boundary fields cancel
each other when collecting all blocks into the whole chain.
This method was used to the describe
the critical line of the XXZ model for $0\leq\Delta\leq1$ and its
basic idea relies on the "restoring" the rotational symmetry of the
model by adding the appropriate boundary terms to the model. The open
chain Hamiltonian is defined as
\begin{widetext}
\begin{eqnarray}
\no H=\frac{J}{4}\sum_{i=1}^{N}h_{i,i+1}~~~,~~~
h_{i,i+1}=\Big[\sigma_{i}^{x}\sigma_{i+1}^{x}+\sigma_{i}^{y}\sigma_{i+1}^{y}+
(\frac{q+q^{-1}}{2})\sigma_{i}^{z}\sigma_{i+1}^{z}
-(\frac{q-q^{-1}}{2})(\sigma_{i}^{z}-\sigma_{i+1}^{z})\Big]~~~,~~
%0<q\leq1 .
\end{eqnarray}
\end{widetext}
Where $q$ is an arbitrary complex number.

To construct a real space RG, we have considered three-site block
(Fig.\ref{3siteblock}) which is important in order to get a
renormalized Hamiltonian of the same form as the original one. In
this case the inter-block ($H^{BB})$ and intra-block ($H^{B}$)
Hamiltonians are:
\begin{widetext}
\begin{eqnarray}
\no
h_{I}^{B}&=&\frac{J}{4}\Big[(\sigma_{1,I}^{x}\sigma_{2,I}^{x}+\sigma_{2,I}^{x}\sigma_{3,I}^{x}+
\sigma_{1,I}^{y}\sigma_{2,I}^{y}+\sigma_{2,I}^{y}\sigma_{3,I}^{y})
+(\frac{q+q^{-1}}{2})(\sigma_{1,I}^{z}\sigma_{2,I}^{z}+\sigma_{2,I}^{z}\sigma_{3,I}^{z})
-(\frac{q-q^{-1}}{2})(\sigma_{1,I}^{z}-\sigma_{3,I}^{z})\Big],\\
\no
H^{BB}&=&\frac{J}{4}\sum_{I=1}^{N/3}\Big[(\sigma_{3,I}^{x}\sigma_{1,I+1}^{x}+
\sigma_{3,I}^{y}\sigma_{1,I+1}^{y})
+(\frac{q+q^{-1}}{2})(\sigma_{3,I}^{z}\sigma_{1,I+1}^{z})
-(\frac{q-q^{-1}}{2})(\sigma_{3,I}^{z}-\sigma_{1,I+1}^{z})\Big].
\end{eqnarray}
\end{widetext}

In this way the correlations between blocks are taken into account
in the coarse graining procedure in terms of the boundary fields of each block.
The ground state of the block
Hamiltonian is doubly degenerate and given by
\begin{widetext}
\begin{eqnarray}
\no
|\psi_{0}\rangle=\frac{1}{\sqrt{2(q+q^{-1}+1)}}(-q^{1/2}|\uparrow\uparrow\downarrow\rangle+
(q^{1/2}+q^{-1/2})|\uparrow\downarrow\uparrow\rangle-q^{-1/2}|\downarrow\uparrow\uparrow\rangle) ,\\
\no
|\psi_{0}'\rangle=\frac{1}{\sqrt{2(q+q^{-1}+1)}}(-q^{1/2}|\uparrow\downarrow\downarrow\rangle+
(q^{1/2}+q^{-1/2})|\downarrow\uparrow\downarrow\rangle-q^{-1/2}|\downarrow\downarrow\uparrow\rangle) .
\end{eqnarray}
\end{widetext}
and the corresponding energy is $e_{0}=-\frac{J}{4}(2+q+q^{-1})$.
%Its details is shown explicitly in \cite{Mdelgado}.We use obtained results
%for studying the renormalization of the entanglement.
The effective Hamiltonian with
renormalized couplings is
\begin{widetext}
\begin{eqnarray}
\no
H^{eff}=\frac{J'}{4}\left[\sum_{i}^{N/3}({\sigma}_{i}^{x}{\sigma}_{i+1}^{x}+{\sigma}_{i}^{y}{\sigma}_{i+1}^{y})+
(\frac{q'+q'^{-1}}{2})({\sigma}_{i}^{z}{\sigma}_{i+1}^{z})
-(\frac{q'-q'^{-1}}{2})(\sigma_{i}^{z}-\sigma_{i+1}^{z})\right],
\end{eqnarray}
\end{widetext}
where
\bea \label{51}
q'=q~~,~~
J'=\xi^{2}(q)J~~,~ ~
\xi(q)=\frac{q+q^{-1}+2}{2(q+q^{-1}+1)} \eea
The remarkable result of eq.(\ref{51}) is that the coupling constant
$q$ when set as a pure phase, or alternatively
$\Delta=(\frac{q+q^{-1}}{2})$ dose not flow under RG transformation
which predicts correctly a line of critical models in the range
$|\Delta|<1$, while $J^{m}$, which is the value of $J$ after $m$ RG
iterations, goes to zero in the limit where $m\rightarrow\infty$, which
in turn defines the scale of energy. This means that in the
critical region the entanglement measure, either concurrence or
entanglement entropy do not evolve through the renormalization of the
anisotropic coupling constant, i.e. it does not rescale the anisotropy parameter
($\Delta'=\Delta$) since every point in this region is a fixed
point.

\begin{figure}
\begin{center}
\includegraphics[width=8cm]{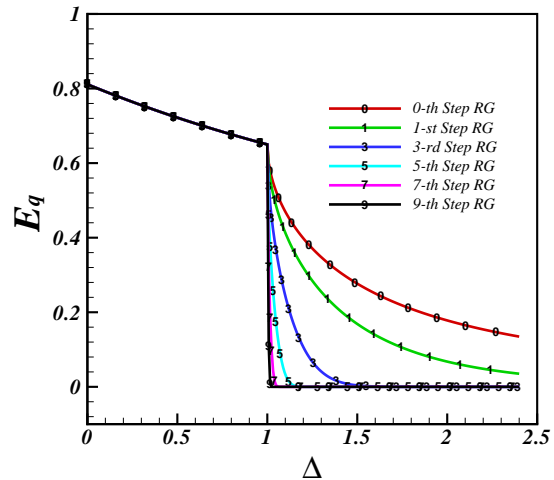}
\caption{(color online) Quantum group manifestation of the evolution
of Entanglement entropy at different RG iterations (steps).}
\label{fig50}
\end{center}
\end{figure}

The density matrix is constructed from either $|\psi_{0}\rangle$ or
$|\psi_{0}'\rangle$. We then sum over site 1 and 3 degrees of
freedom to get the reduced density matrix of site 2 and the rest of
system. The von-Neumann entropy or the entanglement of site 2 and
the whole of system is obtained by the eigenvalues of the reduced
density matrix (similar to Eq.(\ref{eq44})). The entanglement of
site 2 ($E_q$) has been plotted in Fig.\ref{fig50} versus the
anisotropy parameter. For $0 < \Delta\leq 1$ we observe a decrease
of the entanglement versus $\Delta$ which is the effect of
anisotropy to reduce the quantum correlations whilt it is maximum at
$\Delta=0$. Moreover, the reported values for $0 < \Delta\leq 1$ do
not evolve under RG transformation which shows to be the value as
$N\rightarrow \infty$. For $\Delta>1$ we get the evolution of $E_q$
in terms of RG transformation. For higher RG iterations we get the
zero value of entanglement which manifests the uncorrelated nature
of a (classical) Ising state. The nonanalytic behavior which
manifest itself in the first derivative of the entanglement comes as
the critical point $\Delta_{c}=1$ is approached from the Ising
(gapped) phase and is accompanied by a scaling with an exponents like
what were seen before.

%%%%%%%%%%%%%%%%  Fig.quntum group entanglement  %%%%%%%%%%%%%%%%%%%%%%%

%%%%%%%%%%%%%%%%%%%%%%%%%%%%%%%%%%%%%%%%%%%%%%%%%%%%%%%

\section{conclusions \label{conclusion}}

In this work the relation between the entanglement and quantum phase
transition via the renormalization group procedure is addressed. We
have used the idea of renormalization group to study the
quantum information properties of the XXZ chain. The
entanglement entropy and concurrence as two measures of quantum
correlations are used. In order to explore the critical behavior of
the XXZ model the evolution of both entanglement entropy and
concurrence through the renormalization of the lattice were
examined. As the number of RG iterations increases the entanglement
entropy as well as concurrence develops two different values in both
sides of the quantum critical point, where the phase transition
between the spin fluid phase and the Ising like phase occurs. The
phase transition becomes significant which shows a diverging
behavior in the first derivative of the two measures. This
divergence of entanglement measures accompanied by some scaling
behavior near the critical point (as the size of the system becomes
large). The scaling behavior characterizes how the critical point of
the model is touched as the system size is increased. The fact that both measures of
entanglement scale with the same exponent is a significant result of the
large scale behavior of the model near the critical point.
It is
also shown that the nonanalytic behavior of both measures of
entanglement is mirrored from the correlation length exponent in the
vicinity of the critical point. This inspires the behavior of the
entanglement near the critical point is directly connected to the
quantum critical properties of the model.

Compensating the poor results
of the QRG in the critical region of the XXZ model, we
resort to the quantum group method in order to take into account the
correlation between blocks. In this way the massless region is truly
captured and a realistic feature of the evolution of the concurrence
or entanglement entropy is inspired. However, the nonanalytic
behavior and the scaling behavior is the same as before as the
critical point of the model is approached from the gapped phase.

The approach presented here can be used to get the quantum information
properties of quantum systems in an easy way. In other words we get the
properties of a large system dealing with a small block which make it possible
to get analytic results. However, the quantum renormalization group usually
suffers from the poor quantitative results. One should get more accurate results
using the proposed idea with a more complex numerical method like density matrix
renormalization group or exact diagonalzation method.

%%%%%%%%%%%%%%%%%%%%%%%%%%%%%%%%%%%%%%%%%%%%%%%%%%%%%%%%%%%%%%%%%
%The authors would like to thank M. A. Martin-Delgado for fruitful discussions and comments.
\begin{acknowledgments}
This work was supported in part by the Center of Excellence in Complex
Systems and Condensed Matter (www.cscm.ir).
\end{acknowledgments}
%%%%%%%%%%%%%%%%%%%%%%%%%%%%%%%%%%%%%%%%%%%%%%%%%%%%%%%%%%%%%%%%


\section*{References}
\begin{thebibliography}{99}

\bibitem{Bell}
J. S. Bell, Physics \textbf{1}, 195 (1964).

\bibitem{Nielsen}
M. A. Nielsen and I. L. Chuang, \emph{Quantum Computation and
Quantum Communication} (Cambridge University Press, Cambridge,
2000).

\bibitem{Sachdev}
S. Sachdev, \emph{Quantum Phase Transitions} (Cambridge University
Press, Cambridge, 2000).

\bibitem{Osterloh}
A. Osterloh, Luigi Amico, G. Falci and Rosario Fazio, Nature
\textbf{416}, 608 (2002).

\bibitem{Wu}
L.A. Wu, M. S. Sarandy, and D. A. Lidar,Phys. Rev. Lett \textbf{93}
250404 (2004).

\bibitem{latorre}
J. I. Latorre, C. A. L\"utken, E. Rico and G. Vidal, Phys. Rev. A. {\bf 71}, 034301 (2005).

\bibitem{Vidal1}
G. Vidal, J. I. Latorre, E. Rico, and A. Kitaev, Phys. Rev. Lett.
\textbf{90}, 227902 (2003).

\bibitem{Vidal2}
J. Vidal, G. Palacios, and R. Mosseri, Phys. Rev. A \textbf{69},
022107 (2004).

\bibitem{Osborne}
T. J. Osborne and M. A. Nielsen, Phys. Rev. A \textbf{66}, 032110
(2002).

%\bibitem{Latorre}
%J. I. Latorre, E. Rico, and G. Vidal, Quantum Inf. Comput.
%\textbf{4}, 48 (2004).

\bibitem{Bose}
I. Bose and E. Chattopadhyay, Phys. Rev. A \textbf{66}, 062320
(2002).

\bibitem{Verstraete}
F. Verstraete, M. Popp, and J. I. Cirac, Phys. Rev. Lett.
\textbf{92}, 027901 (2004).

%\bibitem{Yang}
%Yang

\bibitem{Zanardi}
P. Zanardi and X. Wang, J. Phys. A \textbf{35}, 7947 (2002)

\bibitem{Gu} Shi-Jian Gu, Shu-Sa Deng, You-Quan Li, and Hai-Qing
Lin,Phys. Rev. Lett. \textbf{93},086402 (2004).

\bibitem{Anfossi}
Alberto Anfossi, Paolo Giorda, and Arianna Montorsi, Phys. Rev.
B \textbf{75}, 165106 (2007)

\bibitem{kargarian}
M. Kargarian, R. Jafari and A. Langari, Phys. Rev. A {\bf 76},
60304 (R) (2007).

\bibitem{fv}
F. Verstraete, J. I. Cirac, J. I. Latorre, E. Rico, and M. M. Wolf
Phys. Rev. Lett. \textbf{94}, 140601 (2005).

\bibitem{pfeuty}
P. Pfeuty, R. Jullian, K.L. Penson in: Real-Space Renormalization,
eds. T.W. Burkhardt, J.M.J. van Leeuwen (Springer, Berlin, 1982) ch.
5.

\bibitem{wilson}
K. G. Wilson, Rev. Mod. Phys. {\bf 47}, 773 (1975).

\bibitem{Shi3}
Shi-Jian Gu, Hai-Qing Lin, and You-Quan Li , Phys. Rev. A
\textbf{68}, 042330 (2003)

\bibitem{Mdelgado}
M. A. Delgado, G. Sierra Phys. Rev. Lett. \textbf{76}, 1146 (1996)

\bibitem{miguel1}
M. A. Martin-Delgado and G. Sierra, Int. J. Mod, Phys. A
\textbf{11}, 3145 (1996).

\bibitem{langari}
A. Langari, Phys. Rev. {\bf B69}, 100402(R) (2004);
A. Langari, Phys. Rev. {\bf B58}, 14467 (1998).

\bibitem{Latorre}
J. I. Latorre, E. Rico, and G. Vidal, Quant.Inf.Comput. \textbf{4},
48 (2004)

\bibitem{jafari}
R. Jafari and A. Langari, Phys. Rev. B {\bf 76}, 014412 (2007);
R. Jafari and A. Langari, Physica {\bf A364}, 213-222 (2006)

\bibitem{Shi1}
Shi-Jian Gu, Guang-Shan Tian, and Hai-Qing Lin, Phys. Rev. A
\textbf{71}, 052322 (2005)

\bibitem{Shi2}
Shi-Jian Gu, Guang-Shan Tian, and Hai-Qing Lin, New Journal of
Physics \textbf{8},61(2006)


\bibitem{entanglement of formation}
W. K. Wootters, Phys. Rev. Lett. \textbf{80}, 2245 (1998).
C. H. Bennett, D. P. DiVincenzo, J. A. Smolin, and W. K. Wooters, Phys. Rev. A, \textbf{54}, 3824 (1996).


\bibitem{Coffman}
V. Coffman, J. Kundu and W.K. Wootters Phys. Rev. A \textbf{61},
052306 (2000)

\bibitem{korepin}
V. E. Korepin, Phys. Rev. Lett \textbf{92},096402 (2004)

\bibitem{Chen}
Y. Chen, P. Zanardi, Z. D. Wang, and F. C. Zhang, New Journal of
Physics 8, 97 (2006), quant-ph/0407228

\bibitem{white}
S.R. White, R.M. Noack, Phys. Rev. Lett. 68, 3487 (1992)

\bibitem{JOsborne}
T. J. Osborne and M. A. Nielsen , Quantum Information Processing
\textbf{1}, 45 (2002). quant-ph/0109024




\end{thebibliography}
\end{document}